\documentclass[11pt,a4paper]{emulateapj}

\usepackage{hyperref}

\usepackage{epsfig}
\usepackage{amsmath}
\usepackage{natbib}
\usepackage{graphicx}
\usepackage{verbatim}
\usepackage{amssymb}

\begin{document}

\title{The fossil record of two-phase galaxy assembly: \\ kinematics and metallicities in the nearest S0 galaxy}

\author{Jacob A. Arnold\altaffilmark{1}}
\author{Aaron J. Romanowsky\altaffilmark{1}}
\author{Jean P. Brodie\altaffilmark{1}}
\author{Laura Chomiuk\altaffilmark{2,3}}
\author{Lee R. Spitler\altaffilmark{4}}
\author{Jay Strader\altaffilmark{2}}
\author{Andrew J. Benson\altaffilmark{5}}
\author{Duncan A. Forbes\altaffilmark{4}}

\altaffiltext{1}{UCO/Lick Observatory, University of California, Santa Cruz, CA 95064, USA}
\altaffiltext{2}{Harvard-Smithsonian Center for Astrophysics, 60 Garden St., Cambridge, MA 02138, USA}
\altaffiltext{3}{National Radio Astronomy Observatory, P. O. Box 0, Socorro, NM 87801, USA}
\altaffiltext{4}{Centre for Astrophysics \& Supercomputing, Swinburne University, Hawthorn, VIC 3122, Australia}
\altaffiltext{5}{California Institute of Technology, MS 105-24, 1200 East California Boulevard, Pasadena, CA 91125}


\begin{abstract}

We present a global analysis of kinematics and metallicity in the nearest S0 galaxy, NGC 3115, along with implications for its assembly history. The data include high-quality wide-field imaging from Suprime-Cam on the Subaru telescope, and multi-slit spectra of the field stars and globular clusters (GCs) obtained using Keck-DEIMOS/LRIS and Magellan-IMACS.  Within two effective radii, the bulge (as traced by the stars and metal-rich GCs) is flattened and rotates rapidly ($v/\sigma \gtrsim$~1.5).  At larger radii, the rotation declines dramatically to $v/\sigma \sim$~0.7, but remains well-aligned with the inner regions.  The radial decrease in characteristic metallicity of both the metal-rich and metal-poor GC subpopulations produces strong gradients with power law slopes of $-0.17\pm 0.04$ and $-0.38\pm 0.06$ dex per dex, respectively. We argue that this pattern is not naturally explained by a binary major merger, but instead by a two-phase assembly process where the inner regions have formed in an early violent, dissipative phase, followed by the protracted growth of the outer parts via minor mergers with typical mass ratios of $\sim$~15--20:1.

\end{abstract}

\maketitle

\begin{figure*}
  \begin{center}
     \epsfxsize=18cm
     \epsfbox{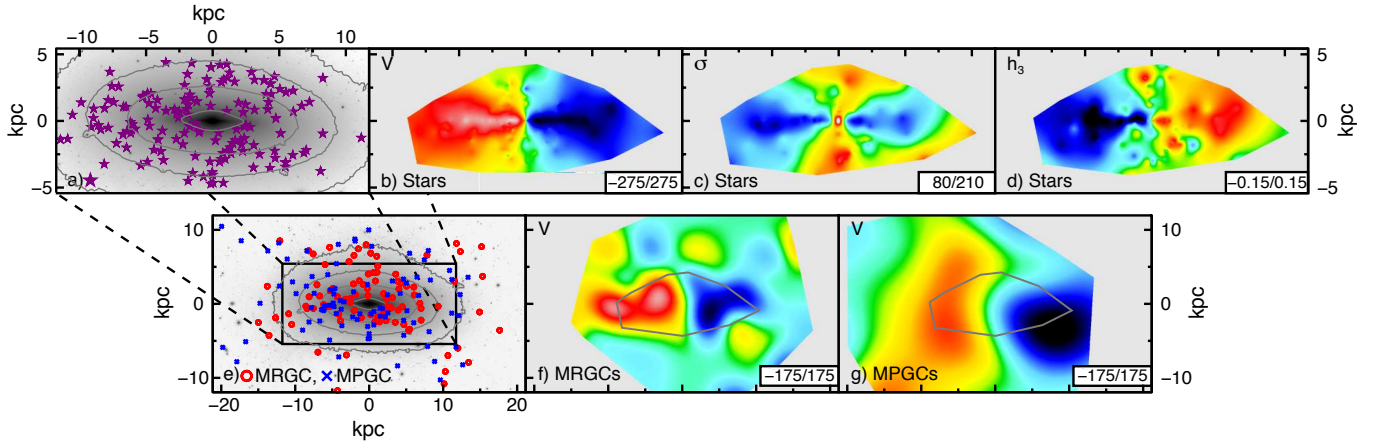}
     \caption{Spatial and kinematic maps of NGC~3115. Top: $i$-band image (a) with SKiMS (stellar light) measurement locations (purple stars). Maps of stellar rotation (b), velocity dispersion (c) and $h_3$ (d), each sharing the same spatial scale as (a). Bottom: Zoomed-out view (e) with spectroscopically confirmed metal-rich (red circles) and metal-poor GC (blue crosses) locations. The inset rectangle denotes the region shown in (a) where selected $i$-band isophotes are also overplotted. Rotation maps, made using the MRGC (f) and MPGC (g) datasets, with the same spatial scale as (e), and overlaid by an outline of the  zoomed-in stellar fields from (b)--(d).
     The kinematic maps were created using kriging techniques for statistically reconstructing sparsely-sampled functions, using ``interpolation'' and ``fitting'' approaches for the stellar and GC datasets, respectively \citep{2007NRC}. The minimum/maximum color-bar scaling is recorded in the bottom right corner of each panel (all in units of km~$\rm{s^{-1}}$ except for (d), which is dimensionless).  
While these maps effectively convey the broad-brush kinematic behavior, we note that interpolating sparse data sets may lead to over-smoothing and anomalous small-scale structure, particularly at larger radius.
     }
     \label{fig:galimage}
  \end{center}
\end{figure*}

\section{Introduction}\label{intro}

The faint outer regions of galaxies are particularly useful probes of the earliest phases of galaxy formation. Here relaxation times are long, and subtle formational clues are preserved not only in their integrated stellar light but also in bright tracer populations such as globular clusters (GCs) and planetary nebulae (PNe).

Recently, wide field observations of  integrated stellar light and PNe \citep{2009MNRAS.398...91P,2009MNRAS.394.1249C} have revealed that the relative kinematic homogeneity among the central regions of early-type galaxies (ETGs) \citep{2007MNRAS.379..401E} gives way to surprising diversity in the outlying parts, e.g., central fast rotators can sometimes rotate slowly at larger radii.

The implication is that inner and outer regions are somewhat decoupled. In the classic gas-rich binary major-merger scenario for forming galaxy bulges, the spatial dependence of dissipational processes could produce dramatic kinematical transitions in radius \citep{2010ApJ...723..818H}.  

Alternatively, inner and outer spheroids may form in two separate phases, which could explain the strong size and internal density evolution of ETGs from $z\sim2$ to today \citep[e.g.][]{2005ApJ...626..680D,2008ApJ...677L...5V}.  This inside-out growth hypothesis means that the inner regions form early as compact stellar spheroids while the outer parts grow later through the accretion of smaller galaxies \citep[e.g.,][]{2009ApJ...699L.178N,2010ApJ...725.2312O}.  

One way to distinguish between the merger and inside-out scenarios is to search for their signatures in the properties of GC systems. In their archetypal study, \citet{1978ApJ...225..357S} used the joint distribution of spatial, kinematical, age, and metallicity properties of GCs to infer the hierarchical buildup of the Milky Way's outer halo.
A key concept here is that building galaxies via the stochastic infall of satellites leads to flattened metallicity gradients at large radius, an effect now observed within the GC systems of two ETGs (\citealt{2009ApJ...703..939H,2011MNRAS.tmp..306F}; cf also \citealt{2010MNRAS.407L..26C}). Likewise, in wet major merger remnants gas dissipation should create metallicity gradients in the central regions, with flatter profiles in the outer regions where any initial gradients have been mixed up in the merger \citep[e.g.,][]{1980MNRAS.191P...1W,2009ApJS..181..135H}. The similar expectations for large-scale metallicity structure from each of these scenarios necessitates additional information, such as from kinematics, to place such results in the proper context.

A useful complication with the GCs is their bimodality, as each galaxy generally has a metal-poor and a metal-rich subpopulation (MPGCs and MRGCs). In the Milky Way these are associated with the stellar halo and the bulge or thick disk, respectively, while in ETGs there are stronger MRGC components reflecting their more dominant bulges. These subpopulations permit the investigation of two distinct phases of galaxy assembly (stellar halo and bulge) far beyond the Local Group.

As part of our ongoing surveys of ETGs 
we have studied the nearest giant S0 galaxy, NGC 3115 \citep[distance 9 Mpc, effective radius $R_e$=57$^{\prime\prime}$=2.5~kpc, $B/T\geq$~0.9, inclination 86$^{\circ}$;][]{1987AJ.....94.1519C}, in unprecedented photometric and spectroscopic detail. In particular, we have obtained extensive kinematic data-sets of its field stars and GCs.  Here we map out the global kinematic and metallicity structure of its dominant, extended bulge, and examine some implications for its assembly history.

In \S \ref{observations} we describe our imaging and spectroscopic observations. Kinematic and metallicity profiles are presented in \S \ref{results}. We discuss how these results constrain the formation mechanisms of the outer bulge and halo in \S~\ref{implications},
and summarize the conclusions in \S~\ref{conclusions}.

\section{Observations}\label{observations}

Here we briefly summarize the observations used in the present analysis; complete details will be presented in a follow-up paper. We obtained $gri$ images with 0.5$^{\prime\prime}$--0.7$^{\prime\prime}$ seeing using Suprime-Cam on the 8.2 m Subaru telescope. Sections of this imaging are shown in Fig. 1a,e to provide orientation for the spectroscopic measurements discussed below.  The photometry allows us to select GC candidates for follow-up spectroscopy and to characterize the distributions of number density and metallicity for the GC system. Color is used as a proxy for metallicity since GCs are generally old \citep[see][]{2006ARA&A..44..193B,2002A&A...395..761K}. The color boundary we use for MPGCs/MRGCs in NGC 3115 is $(g-i)_0$ = 0.91 mag ([Fe/H] = -0.8 dex).

We performed spectroscopy using DEIMOS on the Keck-II telescope to focus on the calcium triplet region \citep[$\sim$8580 $\rm{\AA}$, $R \sim$ 5200; see][]{2009AJ....137.4956R,Foster11}. We have confirmed 150 GCs and also obtained spectra of the background galaxy light in 166 slits for use in a novel technique called ``SKiMS'' \citep[Stellar Kinematics from Multiple Slits:][]{2009MNRAS.398...91P,Foster11}.  Another 15 and 11 unique GC spectra were obtained using LRIS ($\sim$3500--7500 $\rm{\AA}$, $R \sim$ 1100--2200) and IMACS ($\sim$4000--7000 $\rm{\AA}$, $R \sim$ 2200), on the Keck-I and Magellan telescopes, respectively.

GC velocities were measured by cross-correlating their absorption-line spectra with stellar templates.  We imposed a lower velocity cut at 350 km~$\rm{s^{-1}}$ to remove foreground stars.  Fainter SKiMS spectra were coadded to reach a target S/N of 25 using the Voronoi 2D-binning method by \citet{2003MNRAS.342..345C}.  The stellar line-of-sight velocity distribution of each spectrum was parameterized as a truncated Gauss-Hermite series ($V, \sigma, h_3, h_4$) using pPXF \citep{2004PASP..116..138C} and 13 template stars, with parameter uncertainties estimated using Monte Carlo modeling of data-sets with added noise.

We supplement our SKiMS catalog with long-slit stellar kinematics data from \citet[]{2006MNRAS.367..815N} and \citet[]{1982ApJ...256..481I}, and incorporate 20 GC velocities from previous studies \citep{2002A&A...395..761K,2004A&A...415..123P}

\begin{figure*}
  \begin{center}
     \epsfxsize=18cm
     \epsfbox{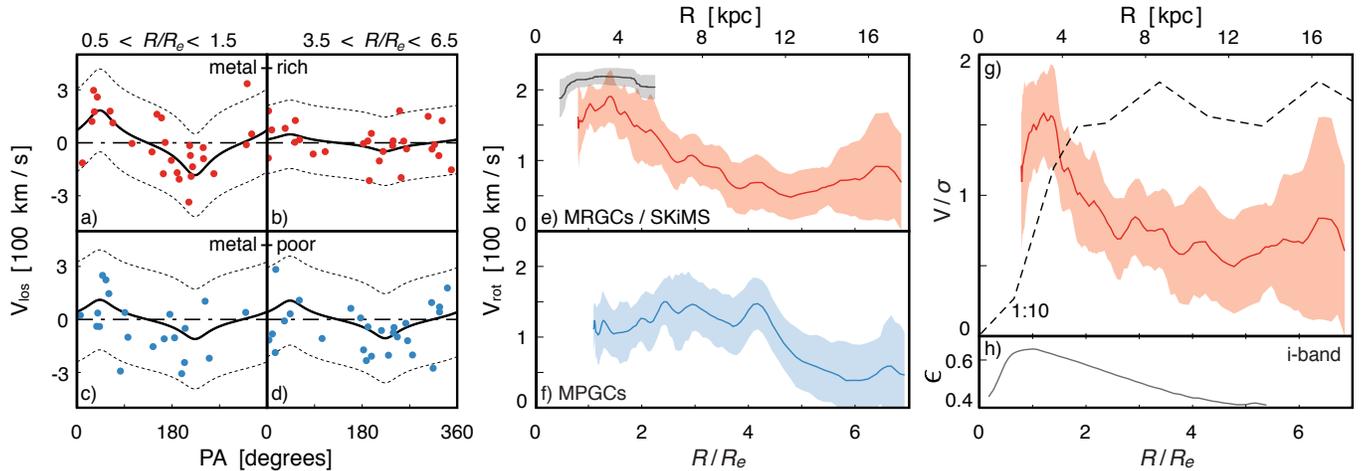}
       \caption{Rotation models for NGC~3115. Panels (a)--(d) plot line-of-sight velocities in two radial bins ($R/R_{\rm e}\approx$ 1, 5) versus position angle for MRGCs (top) and MPGCs (bottom).  Solutions for rotational velocity ($V_{\rm rot}$: solid black curves) and velocity dispersion ($\sigma$) are determined for each dataset using the kinemetry method described in the text (dotted curves denote $V_{\rm rot}$ $\pm$  2$\sigma$).  Smoothed rolling-fit rotational profiles with uncertainty envelopes are shown  for the stellar light (black curve) and MRGCs (red curve) in (e), and for the MPGCs (blue curve) in (f).  In (g) the $v/\sigma$ profile of NGC 3115's MRGC subpopulation (red curve) is compared against a simulated merger remnant with a 1:10 mass-ratio \citep[dashed:][]{2005A&A...437...69B}.  (h) contains the $i$-band ellipticity profile.}
     \label{fig:colpavrot}
  \end{center}
\end{figure*}

\section{Results}\label{results}

Fig. 1 shows two-dimensional maps of the kinematic data. In the top row, the thin disk is evident as a narrow stripe within an extended and rapidly rotating bulge.  The thin disk's lowered velocity dispersion, and anti-correlated $V$ and $h_3$, extend well into the bulge, suggesting an additional embedded component like a very thick disk.  On larger scales (Fig. 1 bottom row), distinct rotation is seen in the MRGC and MPGC subsystems, while there is a general trend for the rotation to weaken in the outer regions.

To simplify the rotation and dispersion trends into one-dimensional profiles, we use a variation of the ``kinemetry'' technique developed for the SAURON survey \citep{2006MNRAS.366..787K} as optimized for data with discrete spatial and velocity sampling \citep{2009MNRAS.398...91P,Foster11}. This method samples the kinematic field (e.g., rotation) using concentric elliptical annuli, and fits the data to flattened sinusoidal models as a function of position angle. For the discrete velocity data (GCs), rotation and dispersion are fitted simultaneously through a maximum likelihood method (Figs. 2a--d). The position angles and ellipticities of the rotation field and the sampling bins ($\rm{PA_{kin}}$, $\rm{\epsilon_{kin}}$) are part of the fit for the SKiMS data but are not well constrained for the discrete velocity data, which we assume follows the stellar isophotes (PA=43.5$^{\circ}$, $\epsilon$=0.5). Our results are insensitive to reasonable variations in these parameters. Uncertainties are estimated via Monte Carlo fitting of mock data-sets.

The resulting rotation profiles for the different subcomponents are shown in Figs. 2e and 2f, where rolling fits with radius are used to capture the details of any radial kinematic transitions (e.g from the inner to outer bulge/halo). Within $\sim$~1.5~$R_{\rm e}$ the MRGC system rotates nearly as rapidly as the stellar bulge, supporting the coevolution of these two components, as also inferred from their similar ages and metallicities \citep{2006MNRAS.367..815N}. At larger radii, this rotation decreases dramatically (see also Figs. 2a,b). The MPGCs have moderate rotation with a decline outside $\sim 4 R_{\rm e}$.  

An alternative rotation profile for the MRGCs is shown in Fig. 2g, after normalizing by the local velocity dispersion. The photometric ellipticity profile is also plotted, showing a decrease with radius that parallels the rotational gradient. The overall implication is for a bulge that has a high degree of rotational flattening in its central regions, while becoming rounder and dispersion-dominated in its outskirts.

Having found kinematic transitions in both GC subpopulations, we look for analogous transitions in the radial metallicity profiles. First we summarize the overall color distribution of the GCs in Fig. 3a, which shows a classic bimodality. We will assume that this bimodality persists with increasing radius, but that the location of the color peaks may shift. At large radii we must cope with the contaminating effects of foreground stars, whose color distribution we also show in Fig. 3a, and use to construct Monte Carlo mock datasets to iteratively correct for the contaminant bias on the color peak locations.

Fig. 3c shows color versus radius, both for individual GC candidates and for the fitted peak locations. Both GC subpopulations have radially-decreasing colors, which we quantify as power-law color gradients with slopes of $-0.05$ and $-0.07$ mag per dex for MPGCs and MRGCs, respectively. Using our own empirical calibration to the $(g-z)$ color used in ACS surveys \citep{2006ApJ...639...95P}, the gradients are $-0.07$ and $-0.10$ mag per dex. Converting to [Fe/H] metallicity \citep{2006ApJ...639...95P}, we estimate gradients of $-0.38\pm 0.06$ and $-0.17\pm 0.04$ dex per dex. 

To our knowledge, this is the first time that metallicity gradients in both GC subpopulations have been measured to large radii in any galaxy besides a few very massive ellipticals (see Section~\ref{intro}). It is also one of the first cases of any galaxy type where joint rotation and metallicity gradients are observed in the halo (see also NGC 4697: \citealt{2005ApJ...627..767M,2009ApJ...691..228M}; and NGC 4125: \citealt{2010A&A...516A...4P}).

\begin{figure*}
  \begin{center}
     \epsfxsize=18cm
     \epsfbox{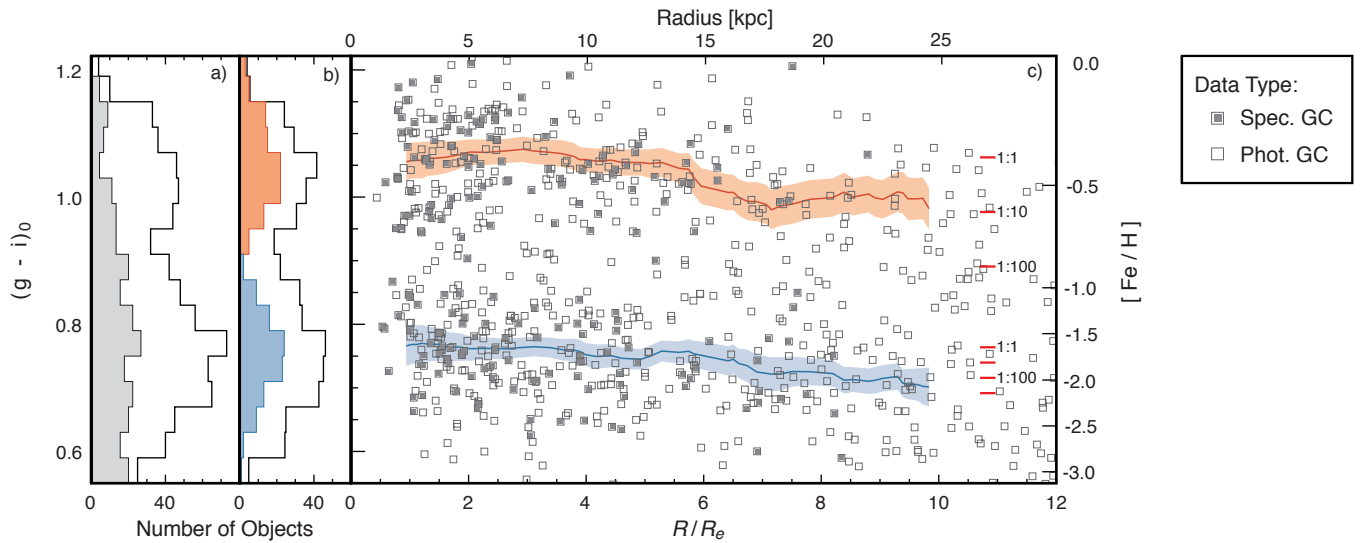}
 \caption{Color distributions $(g-i)_0$ of point sources around NGC~3115.
Left:  Overall photometrically-selected GCs (open histogram)
and of stellar contaminants (filled; as measured at
$R/R_{\rm e} > 14$).
Middle: Resulting true GCs (open) after contaminant
subtraction. Also shown are the spectroscopically confirmed GCs (filled).
Right: Color (left-axis) and [Fe/H] metallicity (right-axis; see text for details) versus radius.
Individual GCs are shown as squares according to the legend at right.
Solid curves denote rolling fits of the peak color of each GC subpopulation 
as estimated from a combination of mixture modeling methods \citep[]{1994AJ....108.2348A} and an iterative Monte Carlo scheme to correct for
the contaminant color-bias.  Associated error
envelopes reflect both statistical and photometric uncertainties.  We convert the
peak color difference between the inner (1--2~$R_e$) and outer ($>$~8~$R_e$) portions of each GC system
into merger mass-ratios (see text), with sample values shown on the right-hand
side. }
     \label{fim:p8}
  \end{center}
\end{figure*}

\section{Implications}\label{implications}

We now consider some possible implications of the rotation and metallicity gradients for NGC 3115's assembly history. The central bulge properties are generally consistent with a standard major merger picture, with the very high amount of rotation in this case indicative of a gas-rich merger with an uneven mass-ratio \citep[e.g.,][]{2005A&A...437...69B,2006MNRAS.372..839N}. Alternatively, the inner bulge might have formed via the inward migration of giant star forming clumps within a turbulent disk fed by cold streams from the cosmic web at early epochs \citep[e.g.,][]{1999ApJ...514...77N,2008ApJ...688...67E,2009ApJ...703..785D}. In either case, the exceptionally high inner-bulge rotation in NGC 3115 may require a residual thick disk component \citep[cf.][]{2001ApJ...554..291C,2010ApJ...722.1666W}. Also, the survival of an old thin disk with a mean age of $\sim$~5--8 Gyr \citep{2006MNRAS.367..815N} means that most of the action in the central regions occurred before $z$~$\sim$~0.5--1, and that there have been no major mergers more recently. 

At larger radii there are generic expectations from major mergers for rotation profile behavior. These remnants are generally expected to have rapid outer rotation resulting from both residual disk spin and the conversion of orbital into internal angular momentum \citep[e.g.,][]{2000MNRAS.316..315B,2001ApJ...554..291C}, in stark contrast to the declining rotation observed in NGC 3115. To make this difference more explicit, we searched through the literature for extended rotation profiles from simulations of major merger remnants, choosing a representative example that comes close to reproducing the central rotation of NGC 3115's bulge and MRGC system. The chosen remnant is the result of a 1:10 spiral-spiral merger from \citet{2005A&A...437...69B}, which we overplot in Fig. 2g, showing the discrepancy in the outer regions between the high rotation predicted and the low rotation observed.

Gas-rich 1:1 merger remnants with small pericenters can also produce declining outer rotation, but the outer ``dry'' part of the remnant is generally expected to show kinematical misalignment with the inner regions \citep[][and in preparation]{2010ApJ...723..818H}. Possible examples of this scenario include NGC 5128 \citep{2004ApJ...602..685P} and NGC 4125 \citep{2010A&A...516A...4P}, but other cases like NGC 3115 with decreasing but well-aligned rotation suggest there must be another explanation (cf. NGC 821 and NGC 3377: \citealp{2009MNRAS.398...91P}; \citealp{2009MNRAS.394.1249C}; and early arguments along these lines by \citealt{1995A&A...293...20S}).

Without exhaustive simulations of major mergers, we cannot rule out the possibility that finely tuned parameters (viewing angle, impact parameter, etc.) would reproduce the observed kinematics of such systems. Nonetheless, it seems more natural to consider a two-phase assembly scenario (Section~\ref{intro}) in which ETGs form inside-out. In this case, inner bulges form at high redshift while subsequent outer bulge and halo growth is driven primarily by dry minor-merger accretion events. The satellites fall in from many different directions and provide little net rotational support \citep{2002ApJ...581..799V,2006MNRAS.365..747A,2007A&A...476.1179B,2010A&A...515A..11Q}. The radial decline in rotation of the MRGC system could then represent a transition from an inner bulge formed in violent, dissipative processes at high redshift, to an outer spheroid (around one third of the bulge mass in the case of NGC 3115) built largely from accreted material over a more protracted period. 

The MPGCs also show a marked rotational decrease, albeit at a larger radius. Here the theoretical picture is less clear, but we postulate an inner metal-poor stellar halo formed in-situ at high $z$ followed by the accretion of outer material that also creates the outer bulge (cf. \citealt{2009ApJ...702.1058Z}).

Declining metallicity profiles are expected in this two-phase assembly scenario since the lower-mass accreted systems should be more metal-poor than the central galaxy \citep{2009ApJ...699L.178N,2009ApJ...697.1290B}.
The prediction (see \S \ref{intro}) is of a downward transition from an inner metallicity profile, whose shape reflects the detailed in-situ star formation history, to an outer profile flattened by radial mixing and primarily composed of accreted material.  We assume this holds for both the stellar and GC metallicity profiles, though systematic offsets may arise from the difference between the ensemble properties of stars and GCs within incoming satellites (e.g. GCs are typically at larger galactocentric radii).
As there are not yet any quantitative predictions, we adopt a schematic model where GC peak metallicities (or colors) are markers of their host galaxies' masses (or luminosities), using the known correlations between these parameters at low $z$ (see below).
The peak GC color in the outer bulge/halo then indicates the characteristic luminosity of the accreted systems.

The color-mass relations may well evolve with time, so we take a more general approach of considering the GC color {\it difference} between the central and outlying regions as an indicator of the characteristic mass-ratio that assembled the outer galaxy. Using the color-mass relations from \citet{2006ApJ...639...95P} for the MPGCs and MRGCs separately, we then generate predicted outer GC peak colors for various stellar mass-ratios 1:$x$ in NGC 3115, overplotting these in Fig. 3c. For the MPGCs we find $x=200$ ($15<x<2900$; the color-mass relation is shallow so this constraint is weak) and for the MRGCs we find $x=8$ (3$<x<$19).  In combination, we estimate $x\sim$~15--20.

Very roughly, the accreted galaxies had luminosities of $\sim 5\times 10^8 L_{B,\odot}$, equivalent to a dwarf elliptical. 
Such galaxies today each host $\sim$~20 MPGCs and $\sim$~3 MRGCs on average \citep{2008ApJ...681..197P}, and could therefore account for a total of $\sim$~350 MPGCs and $\sim$~50 MRGCs in the outer regions of NGC 3115. 
These numbers are different from the $\sim$~260 and $\sim$140 GCs that we estimate NGC 3115 to have outside 2 $R_{\rm e}$, but this discrepancy is not significant given the uncertainties in these calculations. A fundamental complication arises from biased galaxy assembly \citep[][]{2006ARA&A..44..193B}, where the low-mass galaxies that were accreted at high z are thought to have hosted GCs of higher metallicity than their present-day counterparts in lower density environments.  This means that the galaxies accreted by NGC~3115 may have had somewhat lower masses than in our simplified calculation.

As a demonstration of the ongoing accretion process,
NGC~3115 does have a dwarf companion of this luminosity at 45~kpc projected distance, NGC~3115B. 
This galaxy hosts $\sim$~40 GCs \citep{2001AJ....122.1251K} and we estimate the dynamical friction timescale before it disrupts and adds its stars and GCs to the main galaxy to be $\sim~2$~Gyr.

\section{Conclusions}\label{conclusions}

We have found radially-decreasing profiles of rotation and metallicity in the GC system of NGC~3115, and compared these to theoretical expectations. While the central regions were probably formed in violent, dissipative circumstances such as a gas-rich major merger, this event should have produced high or misaligned outer rotation. A more likely scenario for explaining the assembly of the outer bulge and halo is via dry minor mergers and accretion events, whose mass-ratios we have tentatively quantified via the GC metallicities.

Other assembly processes that may contribute to spheroid formation, such as monolithic collapse or quasar-induced expansion (e.g., \citealt{2008ApJ...689L.101F}), have not been considered here and merit future study. Cosmological simulations exploring two-phase assembly predict that up to 80 percent of the stellar mass present in today's early-type galaxies has been accreted \citep{2010ApJ...725.2312O}. Consequently, kinematical and chemical transitions such as those observed here should be ubiquitous.

\acknowledgements
We thank Andi Burkert, Avishai Dekel, Loren Hoffman, Chris Moody, and Joel Primack for useful discussions, the referee Igor Chilingarian for helpful comments and suggestions, and Ewan O'Sullivan for contributing observing time.
Based in part on data collected at Subaru Telescope (operated by the National Astronomical Observatory of Japan), via a Gemini Observatory time exchange (GN-2008A-C-12).
Some of the data presented herein were obtained at the W.~M.~Keck Observatory,
operated as a scientific partnership among the California Institute of Technology, the
University of California and the National Aeronautics and Space Administration, and made possible by the generous financial support of the W.~M.~Keck
Foundation.
This paper includes data gathered with the 6.5 meter Magellan Telescopes located at Las Campanas Observatory, Chile.
This material is based upon work supported by the National Science Foundation under Grants AST-0808099 and AST-0909237 and a Graduate Research Fellowship, and also by the UCSC University Affiliated Research Center's Aligned Research Program. 
AJR was further supported by the FONDAP Center for Astrophysics CONICYT 15010003. 
AJB acknowledges the support of the Gordon and Betty Moore Foundation. 
We acknowledge financial support from the {\it Access to Major Research Facilities Programme}, a component of the {\it International Science Linkages Programme} established under the Australian Government's innovation statement, {\it Backing Australia's Ability}.


\begin{thebibliography}{50}
\expandafter\ifx\csname natexlab\endcsname\relax\def\natexlab#1{#1}\fi

\bibitem[{{Abadi} {et~al.}(2006){Abadi}, {Navarro}, \&
  {Steinmetz}}]{2006MNRAS.365..747A}
{Abadi}, M.~G., {Navarro}, J.~F., \& {Steinmetz}, M. 2006, \mnras, 365, 747

\bibitem[{{Ashman} {et~al.}(1994){Ashman}, {Bird}, \&
  {Zepf}}]{1994AJ....108.2348A}
{Ashman}, K.~M., {Bird}, C.~M., \& {Zepf}, S.~E. 1994, \aj, 108, 2348

\bibitem[{{Bendo} \& {Barnes}(2000)}]{2000MNRAS.316..315B}
{Bendo}, G.~J., \& {Barnes}, J.~E. 2000, \mnras, 316, 315

\bibitem[{{Bezanson} {et~al.}(2009){Bezanson}, {van Dokkum}, {Tal},
  {Marchesini}, {Kriek}, {Franx}, \& {Coppi}}]{2009ApJ...697.1290B}
{Bezanson}, R., {van Dokkum}, P.~G., {Tal}, T., {Marchesini}, D., {Kriek}, M.,
  {Franx}, M., \& {Coppi}, P. 2009, \apj, 697, 1290

\bibitem[{{Bournaud} {et~al.}(2005){Bournaud}, {Jog}, \&
  {Combes}}]{2005A&A...437...69B}
{Bournaud}, F., {Jog}, C.~J., \& {Combes}, F. 2005, \aap, 437, 69

\bibitem[{{Bournaud} {et~al.}(2007){Bournaud}, {Jog}, \&
  {Combes}}]{2007A&A...476.1179B}
---. 2007, \aap, 476, 1179

\bibitem[{{Brodie} \& {Strader}(2006)}]{2006ARA&A..44..193B}
{Brodie}, J.~P., \& {Strader}, J. 2006, \araa, 44, 193

\bibitem[{{Capaccioli} {et~al.}(1987){Capaccioli}, {Held}, \&
  {Nieto}}]{1987AJ.....94.1519C}
{Capaccioli}, M., {Held}, E.~V., \& {Nieto}, J. 1987, \aj, 94, 1519

\bibitem[{{Cappellari} \& {Copin}(2003)}]{2003MNRAS.342..345C}
{Cappellari}, M., \& {Copin}, Y. 2003, \mnras, 342, 345

\bibitem[{{Cappellari} \& {Emsellem}(2004)}]{2004PASP..116..138C}
{Cappellari}, M., \& {Emsellem}, E. 2004, \pasp, 116, 138

\bibitem[{{Coccato} {et~al.}(2010){Coccato}, {Gerhard}, \&
  {Arnaboldi}}]{2010MNRAS.407L..26C}
{Coccato}, L., {Gerhard}, O., \& {Arnaboldi}, M. 2010, \mnras, 407, L26

\bibitem[{{Coccato} {et~al.}(2009){Coccato}, {Gerhard}, {Arnaboldi}, {Das},
  {Douglas}, {Kuijken}, {Merrifield}, {Napolitano}, {Noordermeer},
  {Romanowsky}, {Capaccioli}, {Cortesi}, {de Lorenzi}, \&
  {Freeman}}]{2009MNRAS.394.1249C}
{Coccato}, L., {et~al.} 2009, \mnras, 394, 1249

\bibitem[{{Cretton} {et~al.}(2001){Cretton}, {Naab}, {Rix}, \&
  {Burkert}}]{2001ApJ...554..291C}
{Cretton}, N., {Naab}, T., {Rix}, H., \& {Burkert}, A. 2001, \apj, 554, 291

\bibitem[{{Daddi} {et~al.}(2005){Daddi}, {Renzini}, {Pirzkal}, {Cimatti},
  {Malhotra}, {Stiavelli}, {Xu}, {Pasquali}, {Rhoads}, {Brusa}, {di Serego
  Alighieri}, {Ferguson}, {Koekemoer}, {Moustakas}, {Panagia}, \&
  {Windhorst}}]{2005ApJ...626..680D}
{Daddi}, E., {et~al.} 2005, \apj, 626, 680

\bibitem[{{Dekel} {et~al.}(2009){Dekel}, {Sari}, \&
  {Ceverino}}]{2009ApJ...703..785D}
{Dekel}, A., {Sari}, R., \& {Ceverino}, D. 2009, \apj, 703, 785

\bibitem[{{Elmegreen} {et~al.}(2008){Elmegreen}, {Bournaud}, \&
  {Elmegreen}}]{2008ApJ...688...67E}
{Elmegreen}, B.~G., {Bournaud}, F., \& {Elmegreen}, D.~M. 2008, \apj, 688, 67

\bibitem[{{Emsellem} {et~al.}(2007){Emsellem}, {Cappellari}, {Krajnovi{\'c}},
  {van de Ven}, {Bacon}, {Bureau}, {Davies}, {de Zeeuw}, {Falc{\'o}n-Barroso},
  {Kuntschner}, {McDermid}, {Peletier}, \& {Sarzi}}]{2007MNRAS.379..401E}
{Emsellem}, E., {et~al.} 2007, \mnras, 379, 401

\bibitem[{{Fan} {et~al.}(2008){Fan}, {Lapi}, {De Zotti}, \&
  {Danese}}]{2008ApJ...689L.101F}
{Fan}, L., {Lapi}, A., {De Zotti}, G., \& {Danese}, L. 2008, \apjl, 689, L101


\bibitem[Forbes et al.(2011)]{2011MNRAS.tmp..306F} Forbes, D.~A., Spitler, 
L.~R., Strader, J., Romanowsky, A.~J., Brodie, J.~P., 
\& Foster, C.\ 2011, \mnras, 306 

\bibitem[{{Foster} {et~al.}(2011){Foster}, {Spitler}, {Romanowsky}, {Forbes},
  {Pota}, {Bekki}, {Strader}, {Proctor}, {Arnold}, \& {Brodie}}]{Foster11}
{Foster}, C., {et~al.} 2011, \mnras, submitted

\bibitem[{{Harris}(2009)}]{2009ApJ...703..939H}
{Harris}, W.~E. 2009, \apj, 703, 939

\bibitem[{{Hoffman} {et~al.}(2010){Hoffman}, {Cox}, {Dutta}, \&
  {Hernquist}}]{2010ApJ...723..818H}
{Hoffman}, L., {Cox}, T.~J., {Dutta}, S., \& {Hernquist}, L. 2010, \apj, 723,
  818

\bibitem[{{Hopkins} {et~al.}(2009){Hopkins}, {Cox}, {Dutta}, {Hernquist},
  {Kormendy}, \& {Lauer}}]{2009ApJS..181..135H}
{Hopkins}, P.~F., {Cox}, T.~J., {Dutta}, S.~N., {Hernquist}, L., {Kormendy},
  J., \& {Lauer}, T.~R. 2009, \apjs, 181, 135

\bibitem[{{Illingworth} \& {Schechter}(1982)}]{1982ApJ...256..481I}
{Illingworth}, G., \& {Schechter}, P.~L. 1982, \apj, 256, 481

\bibitem[{{Krajnovi{\'c}} {et~al.}(2006){Krajnovi{\'c}}, {Cappellari}, {de
  Zeeuw}, \& {Copin}}]{2006MNRAS.366..787K}
{Krajnovi{\'c}}, D., {Cappellari}, M., {de Zeeuw}, P.~T., \& {Copin}, Y. 2006,
  \mnras, 366, 787

\bibitem[{{Kundu} \& {Whitmore}(2001)}]{2001AJ....122.1251K}
{Kundu}, A., \& {Whitmore}, B.~C. 2001, \aj, 122, 1251

\bibitem[{{Kuntschner} {et~al.}(2002){Kuntschner}, {Ziegler}, {Sharples},
  {Worthey}, \& {Fricke}}]{2002A&A...395..761K}
{Kuntschner}, H., {Ziegler}, B.~L., {Sharples}, R.~M., {Worthey}, G., \&
  {Fricke}, K.~J. 2002, \aap, 395, 761



\bibitem[{{M{\'e}ndez} {et~al.}(2009){M{\'e}ndez}, {Teodorescu}, {Kudritzki},
  \& {Burkert}}]{2009ApJ...691..228M}
{M{\'e}ndez}, R.~H., {Teodorescu}, A.~M., {Kudritzki}, R., \& {Burkert}, A.
  2009, \apj, 691, 228

\bibitem[{{M{\'e}ndez} {et~al.}(2005){M{\'e}ndez}, {Thomas}, {Saglia},
  {Maraston}, {Kudritzki}, \& {Bender}}]{2005ApJ...627..767M}
{M{\'e}ndez}, R.~H., {Thomas}, D., {Saglia}, R.~P., {Maraston}, C.,
  {Kudritzki}, R.~P., \& {Bender}, R. 2005, \apj, 627, 767

\bibitem[{{Naab} {et~al.}(2006){Naab}, {Jesseit}, \&
  {Burkert}}]{2006MNRAS.372..839N}
{Naab}, T., {Jesseit}, R., \& {Burkert}, A. 2006, \mnras, 372, 839

\bibitem[{{Naab} {et~al.}(2009){Naab}, {Johansson}, \&
  {Ostriker}}]{2009ApJ...699L.178N}
{Naab}, T., {Johansson}, P.~H., \& {Ostriker}, J.~P. 2009, \apjl, 699, L178

\bibitem[{{Noguchi}(1999)}]{1999ApJ...514...77N}
{Noguchi}, M. 1999, \apj, 514, 77

\bibitem[{{Norris} {et~al.}(2006){Norris}, {Sharples}, \&
  {Kuntschner}}]{2006MNRAS.367..815N}
{Norris}, M.~A., {Sharples}, R.~M., \& {Kuntschner}, H. 2006, \mnras, 367, 815

\bibitem[{{Oser} {et~al.}(2010){Oser}, {Ostriker}, {Naab}, {Johansson}, \&
  {Burkert}}]{2010ApJ...725.2312O}
{Oser}, L., {Ostriker}, J.~P., {Naab}, T., {Johansson}, P.~H., \& {Burkert}, A.
  2010, \apj, 725, 2312

\bibitem[{{Peng} {et~al.}(2004){Peng}, {Ford}, \&
  {Freeman}}]{2004ApJ...602..685P}
{Peng}, E.~W., {Ford}, H.~C., \& {Freeman}, K.~C. 2004, \apj, 602, 685

\bibitem[{{Peng} {et~al.}(2006){Peng}, {Jord{\'a}n}, {C{\^o}t{\'e}},
  {Blakeslee}, {Ferrarese}, {Mei}, {West}, {Merritt}, {Milosavljevi{\'c}}, \&
  {Tonry}}]{2006ApJ...639...95P}
{Peng}, E.~W., {et~al.} 2006, \apj, 639, 95

\bibitem[{{Peng} {et~al.}(2008){Peng}, {Jord{\'a}n}, {C{\^o}t{\'e}},
  {Takamiya}, {West}, {Blakeslee}, {Chen}, {Ferrarese}, {Mei}, {Tonry}, \&
  {West}}]{2008ApJ...681..197P}
---. 2008, \apj, 681, 197


\bibitem[Press et al.(2007)]{2007NRC} Press, W.~H., Teukolsky, 
S.~A., Vetterling, W.~T., 
\& Flannery, B.~P.\ 2007, Numerical Recipes 3rd Edition: The Art of Scientific Computing (New York: Cambridge: University Press)


\bibitem[{{Proctor} {et~al.}(2009){Proctor}, {Forbes}, {Romanowsky}, {Brodie},
  {Strader}, {Spolaor}, {Mendel}, \& {Spitler}}]{2009MNRAS.398...91P}
{Proctor}, R.~N., {Forbes}, D.~A., {Romanowsky}, A.~J., {Brodie}, J.~P.,
  {Strader}, J., {Spolaor}, M., {Mendel}, J.~T., \& {Spitler}, L. 2009, \mnras,
  398, 91

\bibitem[{{Pu} {et~al.}(2010){Pu}, {Saglia}, {Fabricius}, {Thomas}, {Bender},
  \& {Han}}]{2010A&A...516A...4P}
{Pu}, S.~B., {Saglia}, R.~P., {Fabricius}, M.~H., {Thomas}, J., {Bender}, R.,
  \& {Han}, Z. 2010, \aap, 516, A4+

\bibitem[{{Puzia} {et~al.}(2004){Puzia}, {Kissler-Patig}, {Thomas}, {Maraston},
  {Saglia}, {Bender}, {Richtler}, {Goudfrooij}, \&
  {Hempel}}]{2004A&A...415..123P}
{Puzia}, T.~H., {et~al.} 2004, \aap, 415, 123

\bibitem[Qu et al.(2010)]{2010A&A...515A..11Q} Qu, Y., Di Matteo, P., Lehnert, M., van Driel, W., \& Jog, C.~J.\ 2010, \aap, 515, A11 

\bibitem[{{Romanowsky} {et~al.}(2009){Romanowsky}, {Strader}, {Spitler},
  {Johnson}, {Brodie}, {Forbes}, \& {Ponman}}]{2009AJ....137.4956R}
{Romanowsky}, A.~J., {Strader}, J., {Spitler}, L.~R., {Johnson}, R., {Brodie},
  J.~P., {Forbes}, D.~A., \& {Ponman}, T. 2009, \aj, 137, 4956

\bibitem[{{Scorza} \& {Bender}(1995)}]{1995A&A...293...20S}
{Scorza}, C., \& {Bender}, R. 1995, \aap, 293, 20

\bibitem[{{Searle} \& {Zinn}(1978)}]{1978ApJ...225..357S}
{Searle}, L., \& {Zinn}, R. 1978, \apj, 225, 357

\bibitem[{{van Dokkum} {et~al.}(2008){van Dokkum}, {Franx}, {Kriek}, {Holden},
  {Illingworth}, {Magee}, {Bouwens}, {Marchesini}, {Quadri}, {Rudnick},
  {Taylor}, \& {Toft}}]{2008ApJ...677L...5V}
{van Dokkum}, P.~G., {et~al.} 2008, \apjl, 677, L5

\bibitem[{{Vitvitska} {et~al.}(2002){Vitvitska}, {Klypin}, {Kravtsov},
  {Wechsler}, {Primack}, \& {Bullock}}]{2002ApJ...581..799V}
{Vitvitska}, M., {Klypin}, A.~A., {Kravtsov}, A.~V., {Wechsler}, R.~H.,
  {Primack}, J.~R., \& {Bullock}, J.~S. 2002, \apj, 581, 799

\bibitem[{{White}(1980)}]{1980MNRAS.191P...1W}
{White}, S.~D.~M. 1980, \mnras, 191, 1P

\bibitem[{{Wuyts} {et~al.}(2010){Wuyts}, {Cox}, {Hayward}, {Franx},
  {Hernquist}, {Hopkins}, {Jonsson}, \& {van Dokkum}}]{2010ApJ...722.1666W}
{Wuyts}, S., {Cox}, T.~J., {Hayward}, C.~C., {Franx}, M., {Hernquist}, L.,
  {Hopkins}, P.~F., {Jonsson}, P., \& {van Dokkum}, P.~G. 2010, \apj, 722, 1666

\bibitem[{{Zolotov} {et~al.}(2009){Zolotov}, {Willman}, {Brooks}, {Governato},
  {Brook}, {Hogg}, {Quinn}, \& {Stinson}}]{2009ApJ...702.1058Z}
{Zolotov}, A., {Willman}, B., {Brooks}, A.~M., {Governato}, F., {Brook}, C.~B.,
  {Hogg}, D.~W., {Quinn}, T., \& {Stinson}, G. 2009, \apj, 702, 1058






\end{thebibliography}
\end{document}